# Asbestiform tremolite within the Holocene late pyroclastic deposits of Colli Albani volcano (Latium, Italy): Occurrence and crystal-chemistry

**Giancarlo Della Ventura • Enrico Caprilli • Fabio Bellatreccia • Arnaldo A. De Benedetti • Annibale Mottana**

**Abstract**  This work relates the occurrence and the characterization of fibrous tremolite within the latest pyroclastic deposits of the Colli Albani (Alban Hills) volcano, to the south-east of Rome (Italy). These mineralizations were observed during a systematic rock-sampling undertaken to complete the geological survey for the new 1:50 000 map of this volcanic area. The examined specimens were collected inside distal deposits correlated to the last Albano Maar activity, which are geographically located within the boundaries of the Nemi community. Tremolite occurs within both carbonate ejecta and

G. Della Ventura (☺)
Dipartimento di Scienze, Università di Roma Tre,
Largo S. Leonardo Murialdo 1, I-00146 ROMA
e-mail: giancarlo.dellaventura@uniroma3.it

the host pyroclastic rocks. It shows up as whitish to light gray coloured aggregates of crystals with fibrous aspect and sericeous brightness. Due to the extremely small crystal dimensions, never exceeding 0.5 μm in diameter, the micro-chemical composition of the fibres could be obtained only by combining P-XRD, SEM-EDX and FTIR methods. Infrared spectroscopy, in particular, proved to be a valuable technique to characterize the studied amphibole. The composition determined is that of a Fe-free F-rich (*c*. 53%) tremolite with significant (*c*. 20%) richterite components in solid-solution. The occurrence of fibrous tremolite in an inhabited place, occurring as natural geological material rather than being due to anthropogenic pollution, should be examined with concern, because it implies complex health and legal responsibilities in the case of mobilization due to extreme climatic events.

**1 Introduction**

During the geological survey aimed at the new geological map 1:50 000 of the Colli Albani (Alban Hills) volcanic area south-east of Rome (Giordano et al., 2010), a systematic sampling of rock specimens was undertaken to complete, with their study, the description of the geological formations and of minerals and rocks occurring in the area. A preliminary optical study of some white samples distinctly discernable among the general greyish or reddish main formations allowed us to observe the presence of diffuse mineralizations consisting of fibrous whitish crystal aggregates. The white aggregates were conspicuous as they occurred as almost monomineralic xenoliths scattered at random within the pyroclastic deposits. More rarely, the pyroclastics themselves showed some additional occurrence of fibrous mineralizations.

Since the presence of asbestiform minerals in rocks is now recognized to be a paramount environmental issue, being a possible source of dangerous particulate matter to the atmosphere (see, e.g., Beneduce et al., 2008 or Gunter et al., 2007a, b for exhaustive treatments of this problem), we decided to perform a detailed characterization of such mineralizations, starting with an evident occurrence that was well-constrained from both the stratigraphic and petrologic viewpoints.

Asbestos amphiboles have been definitively identified as lung carcinogenic fibres (e.g., Gunter et al., 2007a). Related diseases include asbestosis and fibrosis or pleural plaques, which occur after a long and intensive exposure to the fibres, and pleural malignant mesothelioma, which may also occur following acute intense exposure (e.g., de Grisogono and Mottana, 2009; Yao et al., 2010).

The term "asbestos" is used to define several naturally occurring silicates with fibrous shape. Based on the EU 2003/18/EC and Italian DPR 915/82 legislations, six fibrous minerals are regulated as being dangerous for human health. Five of them belong to the amphibole group: actinolite, "amosite", anthophyllite, "crocidolite" and tremolite; the sixth is the serpentine-group mineral chrysotile. It is worth underlining here that "amosite" and "crocidolite" are not mineralogical names: "amosite" is the commercial name (acronym from Asbestos Mines of South Africa) for an amphibole mixture consisting of dominant grunerite, whereas "crocidolite" is a term used for a variety of sodic amphiboles, mainly consisting of riebeckite (cf. Hawthorne et al., 2012 for additional details on the amphibole nomenclature). However, all amphibole minerals display wide ranges of compositions, forming clusters that make them often cross the crystal chemical border lines set up by the current I.M.A. nomenclature, thus posing problems for their classification, as well as troubles for health and legal issues arising from exposure to them (Gunter et al., 2007a; Mazziotti-Tagliani et al., 2008).

Indeed, the presence of fibrous amphiboles in the environment, implies complex health and legal responsibilities, the more so when they occur as natural materials rather than being due to anthropogenic pollution. Thus, we decided to perform a detailed characterization of the observed fibres, the main aim being at assessing exactly their composition and classification, which are the two factors bringing with them the above mentioned legal problems. It must be stressed, however, that besides regulatory considerations, the presence of natural fibres in the environment is anyway a potential source of risk for the population and as such must be characterized.

**2 Geology of the sampled area and occurrence of the studied specimens**

The Colli Albani Volcanic District is a part of the Roman Comagmatic Region (Washington, 1908), also called Latium-Campania Perpotassic Province, which extends along the Tyrrhenian margin of Central Italy. This district (Fig. 1), whose activity lasted from ca. 700 000 years ago to Holocene (Giordano et al., 2006; De Benedetti et al., 2008), is constituted by (from bottom to top): a) the Vulcano Laziale ignimbrite plateau and caldera complex, b) the Tuscolano-Artemisio peri-caldera fissure system and the Faete intra-caldera stratovolcano, and c) the Via dei Laghi maar field (Giordano et al., 2010). The most recent activity is confined to the Via dei Laghi maar field (Giordano et al., 2006), where phreatic to phreatomagmatic eruptions formed (from the oldest to the youngest) the maars of Pantano Secco, Prata Porci, Valle Marciana, Nemi, Ariccia, Laghetto di Giuturna, and Albano (Fig. 1), the latter being the most recent centre of volcanic activity. The young Albano maar centre gave rise to a succession of rock-units which, from the lowermost towards the uppermost one, are the Montagnaccio, Coste dei Laghi, Corona del Lago, Cantone, Peperino Albano, Villa Doria and Albalonga units. The recent geological map 1: 50.000 clarifies well such a complex pile of effusive and explosive products (Giordano et al., 2010).

      Recent studies (Funiciello et al., 2003, De Benedetti et al., 2008) identified in addition a number of lahar deposits due to intermittent water overflows from the Albano

lake, which filled the previous wurmian valley fan around the crater (particularly those toward the north and north-west mountain side) and covered the very last phreatomagmatic deposits from the Albano maar. Such lahars constitute the Tavolato formation, where several individual thin units have been identified (Laurora et al., 2009), the uppermost of which was radiometrically dated 5.8 ka (Giordano et al., 2010). However, stratigraphic and archaeological evidence shows that some of them emplaced even during Roman times (De Benedetti et al., 2008). The lahars of the Tavolato formation represent a chaotic sampling of most if not all eruptive and explosive deposits of the Colli Albani Volcano; thus they deserve particular attention to decipher both the past activity and the present epigenetic formation of minerals, especially for their potential effect on human welfare (cf. Gunter et al., 2007a).

All the recent explosive products and derived lahar deposits are characterized by the presence of xenoliths of sedimentary, metamorphic and magmatic nature. Their study has been used to reconstruct the stratigraphy under the volcanic area using the most varied methods. Referring only to the most recent ones, we point out here with those by Karner et al. (2001), Funiciello et al. (2003), Giordano et al. (2002, 2006), Marra et al. (2003), Soligo et al. (2003), Brigatti et al. (2005), Freda et al. (2006), Gaeta et al. (2006), De Benedetti et al. (2008, 2010), and Laurora et al. (2009). Of particular interest to the present investigation are the studies on the sedimentary carbonate substrate of the volcano, as sampled by the eruption, because carbonate rocks (both lime- and dolostones) represent the bulk material that by reacting gave rise to the fibrous mineralizations. Those studies had been carried out in order to investigate the hydrothermal and geothermal potentials of the volcanic district (Funiciello and Parotto, 1978; Amato and Valensise, 1986; De Benedetti et al., 2010), and yet they contribute also to the knowledge of the environmental situation.

Within the products of the last eruptions of the Via dei Laghi maar field, the xenoliths are concentrated in the breccia deposits at the base of each eruption. In an outcrop near the Nemi soccer stadium, two xenoliths extremely enriched of fibrous minerals have been observed inside distal deposits correlated to the last Albano Maar activity. We studied them in detail.

## 3 Overview of samples and methods

Macroscopically the fibrous minerals appear as whitish to light gray coloured aggregates of crystals with fibrous aspect and sericeous brightness. Several of these aggregates were hand-picked from the host-rock and analyzed using optical and electron microscopy. SEM images were obtained using a Philips XL30 microscope at LIME (Laboratorio interdipartimentale di microscopia elettronica), University Roma Tre. The samples were sputtered with gold to improve the image quality. Figure 2 shows two selected secondary-electron (SE) images. In Figure 2a the mineral occurs as bundles of very thin, soft and wool-looking long crystallites, while Figure 2b, obtained at a greater magnification, shows that a single fibre may be several tenths or hundreds of μm in length and less than 1 μm in diameter.

## 4 X-ray powder diffraction

The X-ray powder pattern (Fig. 3) of the studied sample was collected using a Scintag X1 powder diffractometer equipped with a solid-state X-ray detector, under CuK$\alpha$ Ni-filtered radiation. The pattern shows the analyzed powder to be a virtually monophase amphibole; very minor mica impurities are indicated by the presence of a very weak peak at ~ 9° 2θ (Fig. 3, arrowed). Cell parameters were refined using UnitCell (Holland and Redfern, 1997), starting from literature data for tremolite (Gottschalk et al., 1998). The final unit-cell dimensions are (in Å): a = 9.8242(3); b = 18.0554(6); c = 5.2667(2); β(°) = 105.024(3); V(Å$^3$) = 902.27(4).

## 5 Micro-chemical composition of the fibres

The analysis of fibrous minerals represents a difficult problem that is still poorly resolved. The most common techniques for a proper characterization of fibrous minerals involve the use of X-ray diffraction, X-ray fluorescence, polarized optics or several spectroscopic techniques including FTIR, Raman and Mössbauer. Advantages and disadvantages of these techniques are summarized in Meeker et al. (2003), Gunter et al. (2003, 2007a,b) and Gianfagna et al. (2007). Scanning Electron Microscopy (SEM) is extremely useful for characterising the morphology of mineral particles, and when equipped with an EDS detector may provide semi-quantitative chemical data. However. the quantitative analysis by the EDS systems of particles less than 5 μm in diameter presents many challenges, as recently reviewed by Paoletti et al. (2008). The problems are related to the size and shape of the crystallites, which make it difficult to handle ZAF corrections for mass and X-ray absorption. Efforts to overcome those problems have been described by several authors (e.g. Brigatti et al., 2000; Laskin and Cowin, 2003; Ro et al., 2003; Paoletti et al., 2008), who dealt in particular with the improvement of techniques for EDS analysis of amphibole fibres.

The qualitative standardless EDS spectrum of the fibrous crystals from the Nemi occurrence (Fig. 4) shows intense peaks due to Si, Mg, O and Ca. Minor but well defined peaks due to Na and K are also present; F (CuK$\alpha_1$ peak at 0.68 keV) is barely appreciable. On the basis of these elemental data we can infer a composition close to tremolite; the small but detectable amounts of K and Na, however, suggest departure of its composition toward the richterite composition (Della Ventura et al., 1998, Hawthorne et al., 1996, 1997). The extremely small crystal size prevented a reliable WDS analysis, which would be needed for a definitive chemical characterization of the sample.

## 6 FTIR spectroscopy of the studied fibres

To overcome the analytical problems discussed above, we collected powder FTIR data in the OH-stretching region. FTIR spectroscopy is a powerful method for the analysis of fibrous materials. As a matter of fact, the single fibre can be characterized using a microscope, while powder spectra can be collected with very small amounts of materials (1.0 mg or even less). In the present case, we prepared a KBr disk, mixing approximately 150 mg of KBr with 2 mg of fibres hand-picked from the rock sample and purified under the binocular microscope. The FTIR spectrum was collected with a Nicolet Magna 760 spectrometer, equipped with a globar IR source, a KBr beam splitter and a DTGS detector. Nominal resolution was 4 $cm^{-1}$ and 128 scan were co-added for both sample and background. The resulting spectrum (Fig. 5a) shows a main absorption peak at 3674 $cm^{-1}$, and a well-defined doublet at 3735-3714 $cm^{-1}$.

The band assignment for amphiboles in the richterite-tremolite compositional join is well known from the works of Robert et al. (1989), Della Ventura et al. (1997, 1998, 1999, 2003), Hawthorne et al. (1996, 1997), Gottschalk et al. (1999) and Hawthorne and Della Ventura (2007), mostly done on synthetic compositions.

According to those works, end-member tremolite ($\square Ca_2Mg_5Si_8O_{22}OH_2$, where $\square$ = vacant site) shows a single and very sharp peak in the OH-stretching region centred at 3674 $cm^{-1}$. This band can be assigned to the local configuration M1M1M3-OH-$^A\square$-T1T1 = MgMgMg-OH-$^A\square$-SiSi (notation introduced by Della Ventura et al., 1999). As shown by Gottschalk et al. (1999) synthetic tremolite systematically contains a small amount of cummingtonite component in solid-solution, thus its B-site composition is never equal to $Ca_2$, but it departs toward $Ca_{2-x}Mg_x$, with *x* generally in the range 0.1 to 0.3 apfu. This substitution is detected in the OH spectrum, which may help in quantifying the actual composition at the B-site in tremolite (Gottschalk et al., 1999).

End-member richterite ($^A$NaCaMg$_5$Si$_8$O$_{22}$OH$_2$, where A = Na for richterite and K for potassium richterite) shows a main band at 3735 or 3730 cm$^{-1}$, due to the vibration of an O-H dipole directly bonded to three octahedrally-coordinated Mg cations facing a ring of tetrahedra of composition Si$_6$ and directed toward K or Na at the A site, respectively.

Amphiboles along the join tremolite – potassium-richterite show both bands at 3674 and 3735 cm$^{-1}$, the intensities of which are correlated to the amount of tremolite in solid-solution within richterite (and vice-versa) (Hawthorne et al., 1997). Notably, the FWHM (full width at half maximum) of the tremolite peak increases for intermediate compositions along the join, because of increasing structural disorder due to the distribution of different cations at the available structural sites. However, measurement of the relative intensity of the 3735/3674 cm$^{-1}$ peaks, in this chemically simple system, allows calculation of the empty *vs.* occupied A-site (Hawthorne et al., 1997).

In fluorine-bearing richterite, a new band appears in the OH-spectrum, centered at 3714 or 3710 cm$^{-1}$, respectively, for K and Na at A (Robert et al., 1989). The position of this band is constant for varying amounts of F in the amphibole, whereas its intensity is directly correlated to the amount of fluorine in the sample. This band is assigned to local OH-$^A$(K,Na)-F configurations (Robert et al., 1989). Robert et al. (1999) showed that measurement of the 3735/3714 cm$^{-1}$ intensity ratio provides a reliable quantitative measure of the OH/F composition of the amphibole.

On the basis of the above data, we can confidently conclude that, in agreement with the EDS spectrum given in Figure 2, the fibres examined here have a composition along the tremolite – potassium-richterite join; the presence of (K,Na) at the A-site is balanced by the same amount of a monovalent cation (Na) at the B-site to restore for electroneutrality. The amphibole solid solution also contain some F substituting for OH at the O3 site. The composition of the fibre can be thus expressed as:

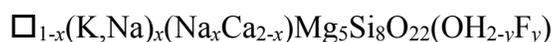
$\square_{1-x}$(K,Na)$_x$(Na$_x$Ca$_{2-x}$)Mg$_5$Si$_8$O$_{22}$(OH$_{2-y}$F$_y$)

Unfortunately the *x* and *y* coefficients cannot be directly determined from the IR spectra because of the simultaneous presence of (K,Na) for □ and F for OH substitutions in the structure, unless some assumptions are made. As shown by Robert et al. (1999), the behaviour of the OH-stretching band of tremolite as a response to the OH-F substitution is significantly different from that of the same band in richterite. In both the richterite-fluororichterite and potassium-richterite – fluoropotassium-richterite series, the OH-stretching band shows a "two-mode behaviour", consistent with a vibrational coupling across the A-site cation (either K or Na). This vibrational coupling in fact generates the double-band pattern visible in Figure 5a at frequency >3700 $cm^{-1}$. In tremolite, where the A-site is vacant, the OH-stretching band simply vanishes as a function of decreasing OH in the amphibole, which is progressively substituted by F.

According to Robert et al. (1999), in potassium-richterite the relative measurement of the 3735-3714 $cm^{-1}$ components provides a measure of the OH/F composition of the "richteritic" part of the studied amphibole; however, the same kind of calculation cannot be done for the "tremolitic" component in solid solution, where, irrespectively of the local OH/F composition, there is only a unique band at 3674 $cm^{-1}$ in the OH-spectrum. Robert et al. (1999) found complete OH-F disorder in richterite for any OH/F ratio; assuming that the same kind of disorder also applies to the tremolite component i.e., assuming that in the tremolite structure there is no preferential OH-OH or F-F local ordering, we can set *y* in the above formula at the same value calculated for the richterite component i.e., to the value obtained from the relative intensities of the 3735/3714 $cm^{-1}$ components. Note that the intensities of the three components in the spectrum (Figure 5a) have been derived from the three Gaussian peaks fitted to the experimental pattern (Fig. 5b). With this method, F = 1.06 and OH = 0.94. We can now calculate *x* (tremolite component) using the relationship: Tr = R/[k + R(1-k)] from Hawthorne et al. (1997), where R = intensity ratio between the (3735 + 3714) and the 3674 $cm^{-1}$ bands, and k = 2.2 (Hawthorne et al., 1997).

Using this indirect method, the estimated chemical composition of the Nemi amphibole turns out to be:

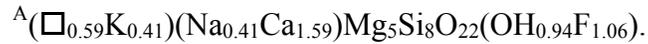

$^A(\square_{0.59}K_{0.41})(Na_{0.41}Ca_{1.59})Mg_5Si_8O_{22}(OH_{0.94}F_{1.06})$.

According to the new classification scheme of the IMA commission for the amphibole nomenclature (e.g., Hawthorne et al., 2012; Hawthorne and Oberti, 2007), the Nemi fibrous amphibole can be regarded as fluoro-tremolite, with significant potassium richterite component.

**7 Conclusions**

Asbestiform minerals in nature are widespread: they occur preferentially in metamorphic rocks of mafic composition. In some areas their concentration may be significant. Notable cases are some outcrops in southern Italy connected with the occurrence of metamorphic rocks such as at Diamante or Amantea (Calabria) (Beneduce et al., 2008), or in several areas on the Alps such as at Ala di Stura (Pacella et al., 2008), Val di Susa (Ballirano et al. (2008), or in the famous chrysotile mine at Balangero (Rossetti and Zucchetti, 1988, Piolatto et al., 1990). Furthermore, there are other occurrences connected with volcanic deposits that are limited in extension, but become important for their potential danger are, such as at Biancavilla (Catania, Sicily). Here, irrespectively of the relatively limited extension, the pathological effects on the population has been well documented (Comba et al., 2003, Cardile et al., 2004).

Increasing anthropic activities connected with landscape modelling, such as excavations for road and building constructions in strongly populated areas where fibrous minerals may potentially occur, favour the possible massive input of airborne particles into the atmosphere. The present work suggests that regulated asbestiform minerals are more common that previously considered in the natural environment, even in areas where they were never detected before and which are close to a large city such as Rome. The geological occurrence of such amphibole mineralizations suggests that the carbonate basement of Colli Albani was involved in metamorphic reaction triggered by fluorine-rich fluids, where the magnesian component of the carbonate rock reacted with silica-rich fluids spreading out from the volcanic magma reservoir at shallow depth and gave rise to amphiboles prior than fluid overpressure made the volcano explode. In other words, the amphibole-rich xenoliths are remnants of the carbonate basement metasomatically reacted with volcanic fluids and brought to the surface by the late explosive activity of the volcano. Lahars then helped spreading these mineralizations all over a limited area at the northern and north-eastern foots of the Alban Hills volcanic system.

The observed occurrence of a potentially dangerous, regulated amphibole species, although being small, urges to try all possible efforts to extend the survey of these mineralization over the entire volcano, while also requesting strongly for a better characterization of the actual contribution of these minerals to the global atmospheric pollution in the Roman area. Indeed, recent reports, such as those arising from the EU-funded Medparticles project (http://www.epidemiologia.lazio.it/medparticles/index.php/en/) definitively demonstrate the link existing between air quality and short-term health effects on population.

**Acknowledgments**: Thanks are due to an anonymous referee and the Rendiconti Lincei Editor for positive suggestions in improving the clarity of the text.

# References

Amato, A. and Valensise, G. (1986) Il basamento sedimentario dell'area albana: risultati di uno studio degli "Ejecta" dei crateri idromagmatici di Albano e Nemi. Mem. Soc. Geol. Ital., 35, 761-767.

Ballirano, P., Andreozzi, G. B. and Belardi, G. (2008) Crystal chemical and structural characterization of fibrous tremolite from Susa Valley, Italy, with comments on potential harmful effects on human health. American Mineralogist, 93, 1349-1355.

Beneduce, P., Di Leo, P., Filizzolla, C., Giano, S.I. and Schiattarella., M. (2008) Valutazione della pericolosità da rilascio di amianto da materiali naturali: un esempio dal Parco Nazionale del Pollino (Italia meridionale). Memorie Descrittive della Carta Geologica d'Italia, 78, 13-30.

Brigatti, M.F., Lugli, C., Cibin, G., Marcelli, A., Giuli, G., Paris, E., Mottana, A. Wu, Z. (2000) Reduction and sorption of chromium by Fe(II)-bearing phyllosilicates: chemical treatments and X-Ray Absorption Spectroscopy (XAS) studies. Clays and Clay Miner, 48, 272-281.

Brigatti, M.F., Caprilli, E., Funiciello, R., Giordano, G., Mottana, A. and Poppi, L. (2005) Crystal chemistry of ferroan phlogopites from the Albano maar lake (Colli Albani volcano, central Italy). Eur. J. Mineral., 17: 611-622.

Cardile, V., Renis, M., Scifo, C., Lombardo, L., Gulino, R., Mancari, B. and Panico, A. M. (2004) Behaviour of new asbestos amphibole fluoro-edenite in different lung cell systems. Int. J. Bioch. Cell Biol., 36, 849-860.

Comba, P., Gianfagna, A. and Paoletti, L. (2003) The pleural mesothelioma cases in Biancavilla are related to the new fluoro-edenite fibrous amphibole. Arch. Environ. Health, 58, 229-232.

De Benedetti, A. A., Funiciello, R., Giordano, G., Caprilli, E., Diano, G. and Paterne, M. (2008) Volcanology, History and Myths of the Lake Albano maar (Colli Albani volcano, Italy). J. Volcanol. Geoth. Res. Special issue "Volcanoes and Human History" (Cashman, K. and Giordano, G. eds.), 176: 387-406.

De Benedetti, A. A., Caprilli, E., Rossetti, F. and Giordano, G. (2010) Metamorphic, metasomatic and intrusive xenoliths of the Colli Albani volcano and their significance for the reconstruction of the volcano plumbing system. *In*: Funiciello, R. & Giordano, G. (eds) The Colli Albani Volcano. Special Publication of IAVCEI, 3. The Geological Society, London, 153-176.

de Grisogono, F.M. and Mottana, A. (2009) The impact of malignant pleural mesothelioma throughout Italy in the years 1995–2002: a geo-referenced study relating death rate to population distribution. Rend. Fis. Acc. Lincei, 20: 117-137.

Della Ventura, G., Robert, J.-L., Raudsepp, M., Hawthorne, F.C. and Welch, M.D. (1997) Site occupancies in synthetic monoclinic amphiboles: Rietveld structure refinement and infrared spectroscopy of (nickel, magnesium, cobalt)-richterite. Am. Mineral., 82, 291-301.

Della Ventura, G., Robert, J.-L. and Hawthorne, F.C. (1998) Characterization of short-range order in potassium-fluor-richterite by infrared spectroscopy in the OH-stretching region. Can. Mineral., 36, 181-186.

Della Ventura, G., Hawthorne, F.C., Robert, J.-L., Delbove, F., Welch, M.D. and Raudsepp, M. (1999) Short-range order of cations in synthetic amphiboles along the richterite - pargasite join. European Journal of Mineralogy, 11, 79-94.

Della Ventura, G., Hawthorne, F.C., Robert, J.-L. and Iezzi, G. (2003) Synthesis and infrared spectroscopy of amphiboles along the tremolite – pargasite join. European Journal of Mineralogy, 15, 341-347.

Freda, C., Gaeta, M., Karner, D.B., Marra, F., Renne, P.R., Taddeucci, J., Scarlato, P., Christensen, J.N. and Dallai, L. (2006) Eruptive history and petrologic evolution of the Albano multiple maar (Alban Hills, Central Italy). Bull. Volcanol. 68, 567-591.

Funiciello, R. and Parotto, M. (1978) Il substrato sedimentario nell'area dei Colli Albani: considerazioni geodinamiche e paleogeografiche sul margine tirrenico dell'Appennino Centrale. Geologica Romana 17, 233-287.

Funiciello, R., Giordano, G. and de Rita, D. (2003) The Albano maar lake (Colli Albani Volcano, Italy): recent volcanic activity and evidence of pre-Roman age catastrophic lahar events. J. Volcanol. Geother. Res. 123, 43-61.

Gaeta, M., Freda, C., Christensen, J.N., Dallai, L., Marra, F., Karner, D.B. and Scarlato, P. (2006) Time-dependent geochemistry of clinopyroxene from the Alban Hills (Central Italy): clues to the source and evolution of ultrapotassic magmas. Lithos 86, 330-246.

Gianfagna, A., Andreozzi, G.B., Ballirano, P., Mazziotti-Tagliani, S. and Bruni, B.M. (2007) Crystal chemistry of the new fibrous fluoro-edenite amphibole of volcanic origin from Biancavilla (Sicily, Italy). Canadian Mineralogist, 45, 249-262.

Giordano, G., de Rita, D., Cas, R.S.A. and Rodani, S. (2002) Valley pond and ignimbrite veneer deposits in the small-volume phreatomagmatic "Peperino Albano" basic ignimbrite, Lago Albano maar, Colli Albani volcano, Italy: influence of topography. J. Volcanol. Geother. Res. 118, 131-144.

Giordano, G., De Benedetti, A.A., Diana, A., Diano, G., Gaudioso, F., Marasco, F., Miceli, M., Mollo, S., Cas, R.A.F. and Funiciello, R. (2006) The Colli Albani caldera (Roma, Italy): stratigraphy, structure and petrology. J. Volcanol. Geother. Res. 155, 49-80.

Giordano, G., Mattei, M. and Funiciello, R. (2010) Geological map of the Colli Albani volcano 1:50 000. I*n*: Funiciello, R. and Giordano, G. (eds) The Colli Albani Volcano. Special Publication of IAVCEI, 3. The Geological Society, London, Insert).

Gottschalk, M., Najorka, J. and Andrut, M. (1998) Structural and compositional characterization of synthetic (Ca,Sr)-tremolite and (Ca,Sr)-diopside solid-solutions. Physics and Chemistry of Minerals, 25, 415-428.

Gottschalk, M., Andrut, M. and Melzer, S. (1999) The determination of cummingtonite content of synthetic tremolite. European Journal of Mineralogy, 11, 967-982.

Gunter, M.E., Dyar, M.D., Twamley, B., Foit, F.F.Jr and Cornelius, S.B. (2003) Composition, Fe3+/∑Fe, and crystal structure of non-asbestiform and asbestiform amphiboles from Libby, Montana. American Mineralogist, 88, 1970-1978.

Gunter, M.. Belluso, E. and Mottana, A. (2007a) Amphiboles: Environmental and health concerns. Reviews in Mineralogy and Geochemistry, 67: 453-516.

Gunter, M.E., Sanchez, M.S. and Williams T.J. (2007b) Characterization of chrysotile samples for the presence of amphiboles, Carey Canadian Deposit, Southeastern Quebec, Canada. Canadian Mineralogist, 45, 263-280.

Hawthorne, F.C. and Della Ventura, G. (2007) Short-range order in amphiboles. Reviews in Mineralogy, 67, 173-222.

Hawthorne. F.C. and Oberti, R. (2007) Classification of the amphiboles. Reviews in Mineralogy and Geochemistry, 67: 55-88.

Hawthorne, F.C., Della Ventura, G. and Robert, J.-L. (1996) Short-range order of (Na,K) and Al in tremolite: An infrared study. American Mineralogist, 81, 782-784.

Hawthorne, F.C., Della Ventura, G., Robert, J.-L., Welch, M.D., Raudsepp, M. and Jenkins, D.M. (1997) A Rietveld and infrared study of synthetic amphiboles along the potassium-richterite - tremolite join. American Mineralogist, 82, 708-716.

Hawthorne, F.C., Oberti, R., Harlow, G.E., Maresch, W.V., Martin, R.F., Schumacher, J.C. and Welch, M.D. (2012) Nomenclature of the amphibole supergroup. American Mineralogist, 97, 2031-2048.

Holland, T.J.B. and Redfern, S.A.T. (1997) Unit cell refinement from powder diffraction data: the use of regression diagnostic. Min. Mag., 61, 65-77.

Karner, D.B., Marra, F. and Renne, P.R. (2001) The history of the Monti Sabatini and Alban Hills volcanoes: groundwork for assessing volcano-tectonic hazard for Rome. J. Volcanol. Geother. Res. 107, 185-219.

Laskin, A. and Cowin, J.P. (2003) Automated single-particle SEM/EDX analysis of submicrometer particles down 0.1 μm. Analytical Chemistry, 73, 1023-1029.

Laurora, A., Malferrari, D., Brigatti, M.F., Mottana, A., Caprilli, E., Giordano, G. and Funiciello, R. (2009) Crystal chemistry of trioctahedral micas in the top sequences of the ColliAlbani volcano, Roman Region, central Italy. Lithos, 113: 507-520.

Marra, F., Freda, C., Scarlato, P., Taddeucci, J., Karner, D.B., Renne, P.R., Gaeta, M., Palladino, D.M., Trigila, R. and Cavarretta, G. (2003) Post Caldera activity in the Alban Hills Volcanic district (Italy): $^{40}Ar/^{39}Ar$ geochronology and insights into the magma evolution. Bull. Volcanol. 65, 227-247.

Mazziotti-Tagliani, S., Andreozzi, G.B., Bruni, B.M., Gianfagna, A., Pacella A. and Paoletti, L. (2008) Quantitative chemistry and compositional variability of fluorine fibrous amphiboles from Biancavilla (Sicily, Italy). Per. Mineral., 78, 65-74.

Meeker, G.P., Bern, A.M., Brownfield, I.K., Lowers, H.A., Sutley, S.J., Hoefen, T.M. and Vance, J.S. (2003) The composition and morphology of amphiboles from the Rainy Creek complex, near Libby, Montana. AmericanMineralogist, 88, 1955-1969.

Pacella, A., Andreozzi, G.B., Ballirano, P. and Gianfagna, A. (2008) Crystal chemical and structural characterization of fibrous tremolite from Ala di Stura (Lanzo valley, Italy). Per. Mineral., 77, 51-62.

Paoletti, L., Bruni, B.M., Arrizza, L., Mazziotti-Tagliani, S. and Pacella, A. (2008) A micro-analytical SEM-EDS method applied to the quantitative chemical composition of fibrous amphiboles. Periodico di Mineralogia, 77, 63-73.

Piolatto, G., Negri, E., La Vecchia, C., Pira, E., Decarli, A. and Peto J. (1990) An update of cancer mortality among chrysotile asbestos miners in Balangero, northern Italy. British Journal of Industrial Medicine, 47, 810-814.

Ro, C., Osan, J., Szaloki, I., de Hoog, J., Worobiec, A. and Van Grieken, R. (2003) A Monte Carlo program for quantitative electron-Induced X-ray analysis of individual particles. Analytical Chemistry, 75, 851-859.

Robert, J.L., Della Ventura, G. and Thauvin, J.L. (1989) The infrared OH stretching region of synthetic richterites in the system $Na_2O$ $K_2O$ $CaO$ $MgO$ $SiO_2$ $H_2O$-HF. European Journal of Mineralogy, 1, 203 211.

Robert, J.-L., Della Ventura, G. and Hawthorne, F.C. (1999) Near-infrared study of short-range disorder of OH and F in monoclinic amphiboles. American Mineralogist, 84, 86-91.

Rossetti, P. and Zucchetti, S. (1988) Early-alpine ore parageneses in the serpentinites from the Balangero asbestos mine and Lanzo Massif (Internal Western Alps). Rendiconti SIMP, 43, 139-149.

Soligo M., Tuccimei P., Giordano G., Funicello R. and de Rita D. (2003) U-series dating of a carbonate level underlying the peperino Albano phreatomagmatic ignimbrite (Colli Albani, Italy). Quaternario, 16(bis) 115-120

Washington, H.S. (1908) The Roman comagmatic region. Carnegie Insitution Washington, publ. 57, 199 pp.

Yao, S., Della Ventura, G. and Petibois, C. (2010) Analytical characterization of cell-asbestos fibers interactions in lung pathogenesis. Analytical and Bioanalytical Chemistry, 397, 2079-2089.

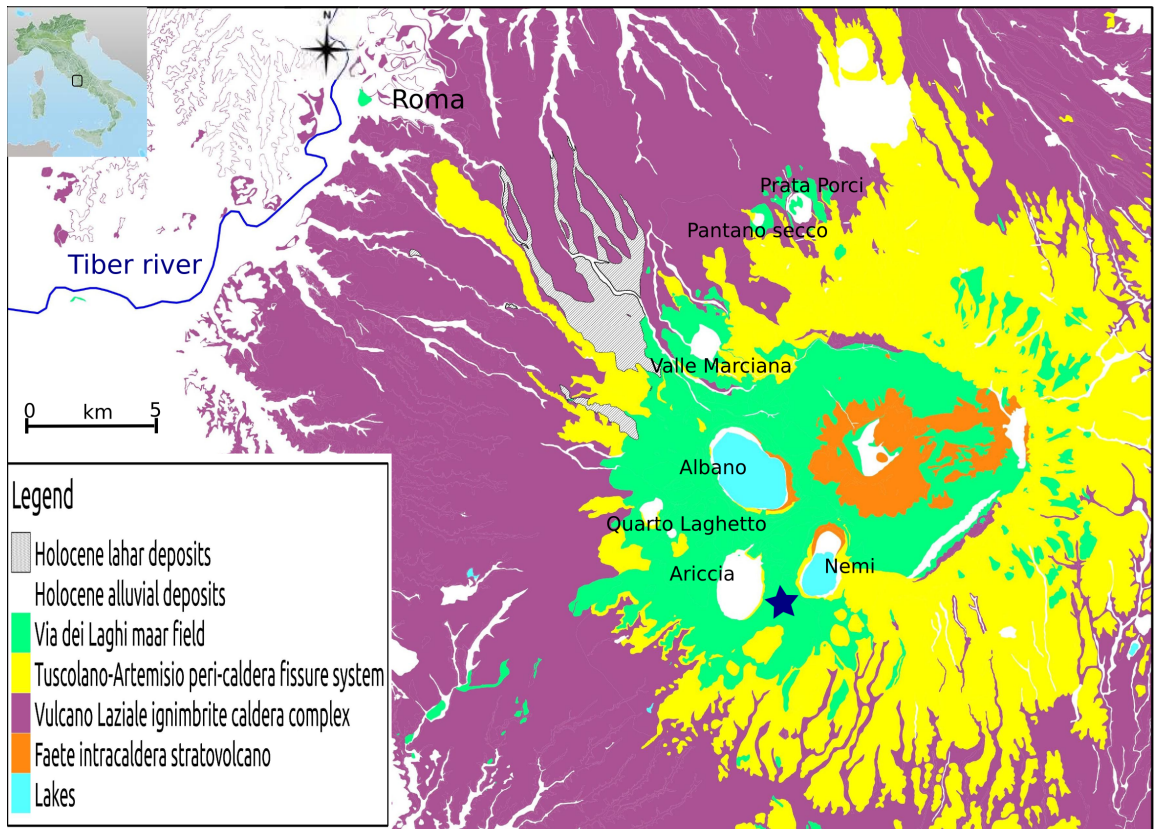

**Fig. 1** Geological sketch map of Colli Albani volcano (from Giordano et al., 2010). The sample location is indicated by a star.

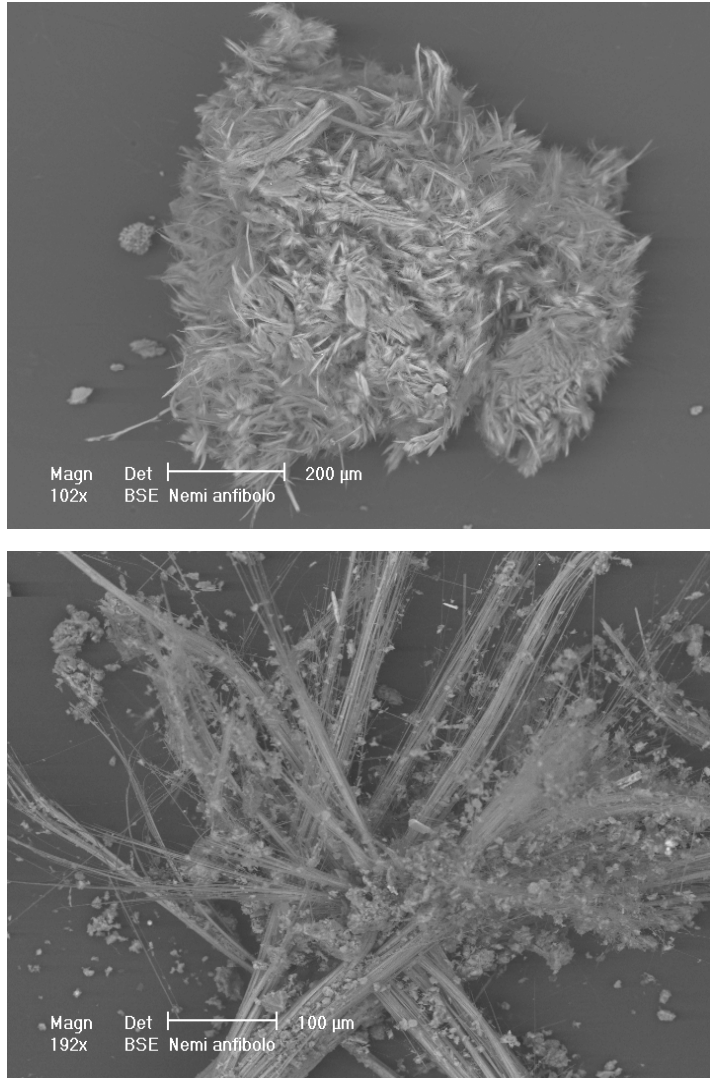

**Fig. 2** SEM-EDS images of the studied sample from Nemi

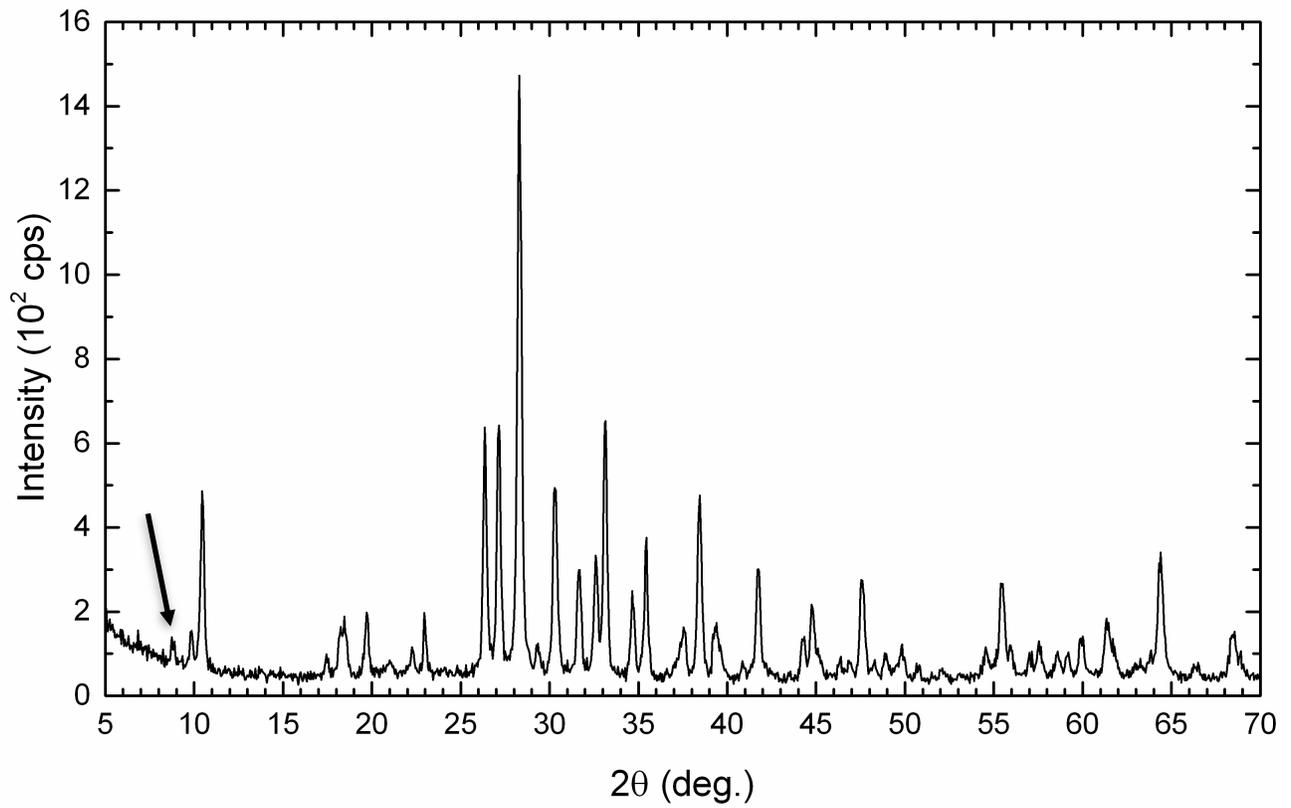

**Fig. 3** XRD pattern of the studied sample from Nemi; Cu Kα Ni-filtered radiation; the mica (001) peak is indicated by an arrow.

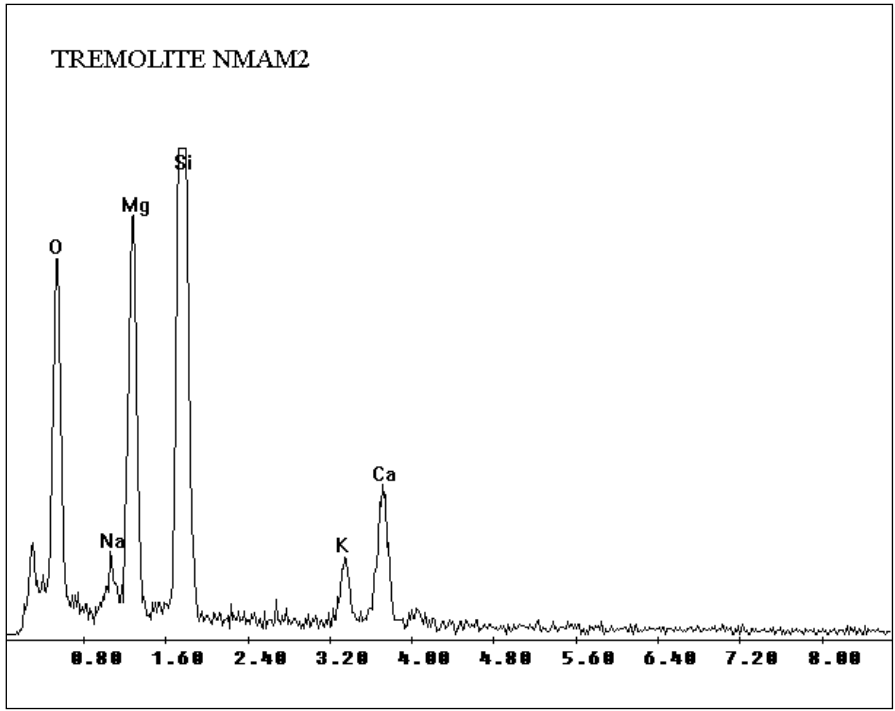

**Fig. 4** EDS fluorescence spectrum of the examined fibrous sample from Nemi

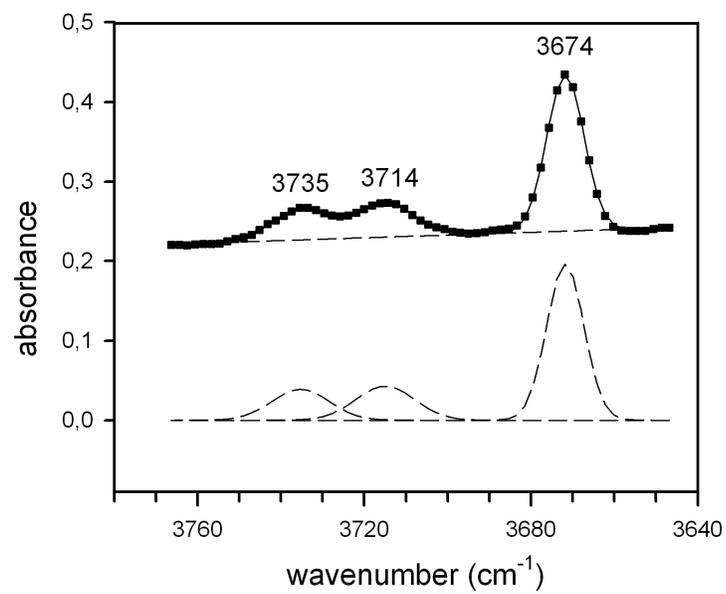

**Fig. 5** (a) FTIR powder spectrum in the principal OH-stretching region; (b) the same spectrum as in (a) decomposed using Gaussian components